\documentclass[12pt,fleqn]{article}
\usepackage{amsmath}

\usepackage{graphicx}
\usepackage{pdftricks}
\usepackage{lineno}
\usepackage{dcolumn}
\RequirePackage{lineno}
\usepackage{epstopdf}

\begin{document}
\title{\Large Thermodynamic calculation of spin scaling functions}

\author{
 George Ruppeiner\footnote{ruppeiner@ncf.edu}\\
 Division of Natural Sciences\\
 New College of Florida\\
 5800 Bay Shore Road\\
 Sarasota, Florida 34243}

\maketitle

\begin{abstract}
Critical phenomena theory centers on the scaled thermodynamic potential per spin $\phi(\beta, h)=|t|^{p}Y(h|t|^{-q})$, with inverse temperature $\beta=1/T$, $h=-\beta H$, ordering field $H$, reduced temperature $t=t(\beta)$, critical exponents $p$ and $q$, and function $Y(z)$ of $z=h|t|^{-q}$. I discuss calculating $Y(z)$ with the information geometry of thermodynamics. Scaled solutions obtain with three admissible functions $t(\beta)$: 1) $t=e^{-J\beta}$, 2) $t=\beta^{-1}$, and 3) $t=\beta_C-\beta$, where $J$ and $\beta_C$ are constants. For $p=q$, information geometry yields $Y(z)=\sqrt{1+z^2}$, consistent with the one-dimensional (1D) ferromagnetic Ising model.
\end{abstract}

\noindent Keywords: Information geometry of thermodynamics; Scaled equation of state; Ising model; Spherical model; Thermodynamic curvature; Phase transitions and critical points.
\\

\par
Critical point theory deals with systems with long-range order, measured by a diverging correlation length $\xi$ \cite{Stanley1999, Pathria2011}. Here, it is hard to evaluate the partition function $Z$ by summing over microstates. In this letter, I approach the problem of long-range order with a method not based on an explicit summation over microstates. I combine the curvature scalar $R$ from the information geometry of thermodynamics with hyperscaling, to get a differential equation for the thermodynamics.

\section{Introduction}

\par
I employ the language of magnetic systems, for which the thermodynamic potential per spin is $\phi=\phi(\beta,h)$, where $\beta=1/T$, with $T$ the temperature, and $h=-\beta H$, with $H$ the ordering field. Boltzmann's constant $k_B=1$. In the language of Pathria and Beale \cite{Pathria2011}, $\phi$ is the $q$-potential per spin: $\phi(\beta,h)=\mbox{ln}Z/N$, and $N$ is the number of spins $(N\to\infty)$. The heat capacity per spin at constant $H$ is $C_H = T(\partial s/\partial T)_H$, with entropy per spin $s = \phi-\beta \phi_{,\beta}-h \phi_{,h}$. The magnetic susceptibility is $\chi_T = (\partial m/\partial H)_T$, with magnetization per spin $m = -\phi_{,h}$. The comma notation denotes differentiation.

\par
Critical phenomena theory centers on scaling and universality, expressed by a scaled relation: 

\begin{equation}\phi(\beta, h)=n_1 |t|^{p}\,Y(n_2\,h|t|^{-q}),\label{10}\end{equation}

\noindent where $p$ and $q$ are critical exponents, $t=t(\beta)$ is the reduced temperature, $n_1$ and $n_2$ are scaling constants, and $Y(z)$ is a universal function of

\begin{equation}z=h |t|^{-q}.\label{20}\end{equation}

\noindent I consider critical points both at a finite critical temperature $T_C$, and at zero temperature.

\par
\begin{figure}
\begin{minipage}[b]{0.5\linewidth}
\includegraphics[width=2.0in]{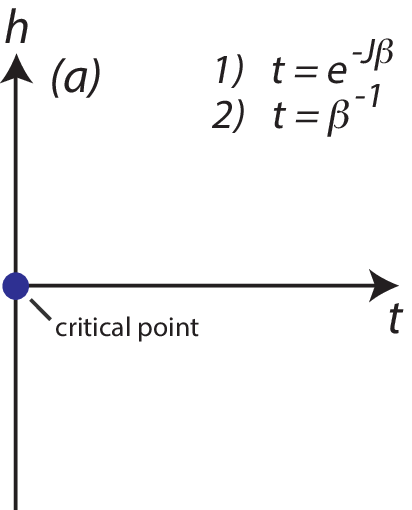}
\end{minipage}
\hspace{0.2 cm}
\begin{minipage}[b]{0.5\linewidth}
\includegraphics[width=2.0in]{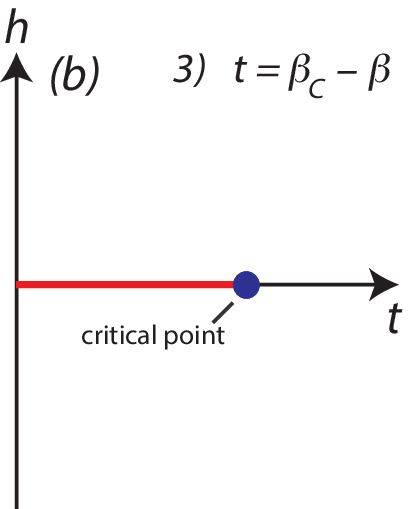}
\end{minipage}
\caption{The critical region in $(t,h)$ space: a) critical temperature $T_C=0$, with no first-order phase transition line, and b) $T_C>0$, with a first-order phase transition line shown in red.}
\label{fig:1}
\end{figure}

As I will show explicitly, the information geometric method allows three admissible functions leading to scaled solutions for $t=t(\beta)$:

\begin{equation}1)\,t=e^{-J\beta}, 2)\,t=\beta^{-1},\mbox{ and }3)\,t=\beta_C-\beta,\label{30}\end{equation}

\noindent where $J$ is a constant and $\beta_C=1/T_C$. These three admissible functions for $t=t(\beta)$ are familiar from the modern theory of critical phenomena. Figure \ref{fig:1} sketches the corresponding critical point scenarios. The function 3) was identified previously as admissible in the information geometric approach \cite{Ruppeiner1998}. The contribution in this letter is the identification of the exponential function 1) and the power law function 2) as admissible as well. These two functional forms are characteristic of critical points at zero temperature.

\par
The information geometric method allows for the determination of $Y(z)$, based on the character of the critical point, and the values of the critical exponents. This is difficult to accomplish with statistical mechanics, and very few cases have been worked out.

\section{The 1D Ising model}

\par
A worked example is the one-dimensional (1D) ferromagnetic Ising model consisting of a chain of $N$ spins ($N\to\infty$) enumerated by an index $i$. The $i$'th spin has state $\sigma_i=+1/-1$, corresponding to spin up/down. The Hamiltonian is

\begin{equation}\displaystyle\mathcal{H} = -J\sum_{i=1}^N\sigma_i\sigma_{i+1} - H\sum_{i=1}^N\sigma_{i},\label{40}\end{equation}

\noindent where $J>0$ is the coupling constant, and we assume periodic boundary conditions: $\sigma_{N+1}=\sigma_1$.

\par
Evaluating the partition function with statistical mechanics leads to \cite{Pathria2011}:

\begin{equation} \phi(\beta,h)=\mbox{ln}\left(e^{J\beta}\cosh h +\sqrt{e^{2J\beta}\,\mbox{sinh}^2 h + e^{-2J\beta}}\,\right), \label{50}\end{equation}

\noindent yielding a critical point at $h=0$ and $T=0$, where $\chi_T$ and $\lvert R\rvert$ diverge. Nelson and Fisher \cite{Nelson1975} worked out the scaled form for the singular part of Eq. (\ref{50}) near the critical point:

\begin{equation} \phi=e^{-2 J\beta}\,Y(h\,e^{2 J\beta}), \label{60}\end{equation}

\noindent with

\begin{equation} z=h\,e^{2 J\beta}, \label{70}\end{equation}

\noindent and

\begin{equation}Y(z)=\sqrt{1+z^2}.\label{80}\end{equation}

\noindent These authors took $t(\beta)=e^{-J\beta}$, and $p=q=2$.

\par
The critical exponents of this model are ambiguous since a dimensionless factor may be removed from $J$ in $t=e^{-J\beta}$, and put into $p$ and $q$. But this will pose no difficulties for the information geometric method since all exponent choices lead to the same scaled equation of state if $p=q$.

\section{Information geometry of thermodynamics}

\par
Strong features of the information geometric calculation method are: 1) the scaled universal function in Eq. (\ref{10}) fits naturally into the structure, 2) $Y(z)$ may be calculated by solving an ordinary differential equation, and 3) strong constraints are put on appropriate choices for $t=t(\beta)$. We also have a special result for the case $p=q$, a case that includes the 1D ferromagnetic Ising model.

\par
Essential in information geometry is the invariant thermodynamic curvature $R$, resulting from the thermodynamic metric elements $g_{\alpha\beta} = \phi_{,\alpha\beta}$ \cite{Ruppeiner1995}. Here, the thermodynamic coordinates are $(x^1, x^2) = (\beta, h)$. For spin systems, $R$ has units of lattice constants to the power of the spatial dimension \cite{Ruppeiner2015a}. For weakly interacting systems, $\lvert R\rvert$ is small, and on approaching a critical point, $R$ diverges to negative infinity. Here, I use the sign convention of Weinberg \cite{Weinberg1972}, where the curvature of the two-sphere is negative.

\par
It has long been argued \cite{Ruppeiner1995, Ruppeiner1979, Johnston2003} that near a critical point

\begin{equation} \lvert R\rvert\propto\xi^d, \label{100}\end{equation}

\noindent where $d$ is the spatial dimension. In addition, hyperscaling from the theory of critical phenomena has \cite{Widom1974, Goodstein1975}:

\begin{equation} \lvert \phi \rvert\propto\xi^{-d}. \label{110}\end{equation}

\noindent Eliminating $\xi^d$ between these two proportionalities leads to

\begin{equation} R=-\frac{\kappa}{\phi}, \label{120}\end{equation}

\noindent where $\kappa$ is a dimensionless constant of order unity that the solution method determines. The known expression of $R$ for spin systems allows us to write Eq. (\ref{120}) as \cite{Ruppeiner1995}

\begin{equation}\phi\left| \begin{array}{ccc} \phi_{,11}& \phi_{,12}& \phi_{,22}\\ \phi_{,111}&\phi_{,112}&\phi_{,122}\\ \phi_{,112}&\phi_{,122}&\phi_{,222} \end{array}\right| \displaystyle =-2\kappa\left| \begin{array}{cc} \phi_{,11} & \phi_{,12}\\\phi_{,12}&\phi_{,22}\end{array}\right|^2.\label{130}\end{equation}

\par
This geometric equation is a third-order partial differential equation (PDE) for $\phi$. It is written in terms of the macroscopic thermodynamic parameters, with its mesoscopic roots entirely hidden. Although the argument above for this equation is somewhat loose, I conjecture that Eq. (\ref{130}) is exact near the critical point. This equation has seen success in varied scenarios, not all connected with critical phenomena; see Table \ref{tab:table1}.

\begin{table}[h!]
\caption{Tests of the geometric equation. $n$ denotes the number of independent thermodynamic variables, $d$ the spatial dimension, $a\#$ the number of analytic sections, and $f$ the number of fit parameters for the scaled equation.}
\label{tab:table1}
\centering
\begin{tabular}{lccccl}\\
\hline
\hline
System										& $n$	& $d$	& $a\#$	& $f$		& notes			\\
\hline
mean field theory \cite{Ruppeiner1991}				& $2$	& $-$	& $1$	& $0$	& exact			\\
critical point \cite{Ruppeiner1991}					& $2$	& $3$	& $2$	& $2$	& $\chi^2\sim 1$	\\
galaxy clustering \cite{Ruppeiner1996}				& $2$	& $3$	& $1$	& $0$	& qualitative		\\
corrections to scaling \cite{Ruppeiner1998}			& $3$	& $3$	& $2$	& $3$	& unclear			\\
ideal gas paramagnet \cite{Kaviani1999}				& $3$	& $3$	& $1$	& $0$	& exact			\\
power law interacting fluids \cite{Ruppeiner2005}		& $2$	& $3$	& $1$	& $1$	& qualitative		\\
unitary fermi fluid \cite{Ruppeiner2014, Ruppeiner2015b}	& $2$	& $3$	& $2$	& $4$	& $\chi^2\sim 2$	\\
black holes \cite{Ruppeiner2018}					& $2$	& $-$	& $1$	& $1$	& unclear			\\
ferromagnetic Ising							& $2$	&$1$		& $1$	& $0$	& exact			\\
\hline
\hline
\end{tabular}
\end{table}

\section{Results}

\par
For each of the three functions for $t(\beta)$ in Eq. (\ref{30}), the PDE Eq. (\ref{130}) simplifies to a third-order ordinary differential equation (ODE) for $Y(z)$ on substituting the scaled expression in Eq. (\ref{10}) for $\phi$. To see the reason for this reduction in complexity, imagine expanding out the two determinants in Eq. (\ref{130}). The result is an equation consisting of nine terms. Each term contains a quadruple of four factors of $\phi$, four derivatives with respect to $\beta$, and four derivatives with respect to $h$. For each of the three functional forms in Eq. (\ref{30}), differentiating $\phi$ or its derivatives with respect to $\beta$ or with respect to $h$ pulls out a factor of a power of $t(\beta)$. These common factors in the quadruples all cancel out, leaving only $z$, $Y(z)$, and its derivatives. I found no other functions $t(\beta)$ that result in this cancellation of factors of $t(\beta)$.

\par
Including the scaling constants $n_1$ and $n_2$ in $\phi$, as in Eq. (\ref{10}), likewise results in a cancellation of all factors $n_1$ and $n_2$. The same ODE, with the same solution $Y(z)$, results. The ODE and $Y(z)$ generally depend only on the critical exponents $p$ and $q$. This dependence is expected from the theory of critical phenomena, where changing the critical exponents put us into a different universality class, with different $Y(z)$'s.

\par
Consider now in detail the solution with the exponential form for $t(\beta)$:

\begin{equation}\phi=\exp(-p J\beta)\,Y(h\exp[q J\beta]).\label{135}\end{equation}

\noindent This form has not been previously considered in the context of information geometry. Particularly simple is the case $q=p$, of which the 1D ferromagnetic Ising model is an example. For $q=p$, Eq. (\ref{130}) simplifies to

\begin{equation} \frac{Y \left(z Y Y^{(3)}+z^2 Y''^2+2 Y Y''-z^2 Y' Y^{(3)}-2 z Y' Y''\right)}{2 \left(Y-z Y'\right)^2 Y''}=\kappa, \label{140}\end{equation}

\noindent independent of the value of $p$. The cancellation of $p$ is physically necessary because we may pull any factor out of $J$ and put it into $p$ without changing the physics, as I remarked in connection with the Ising model. More generally, I add that for models with $t=\exp(-J\beta)$ and $q=n p$, with $n$ a constant factor, $p$ cancels out as well, leaving $Y(z)$ depending only on $n$.

\par
To solve Eq. (\ref{140}), start by assuming that $Y(z)$ is analytic at $z=0$, an assumption generally made in theories of critical phenomena for $\beta<\beta_C$ \cite{Griffiths1967}, and certainly the case in Eq. (\ref{80}). Also assume that $Y(z)$ is symmetric about $z=0$, a general feature of the basic Ising spin models. We have the series

\begin{equation} Y(z) = y_0 + y_2 z^2 + y_4 z^4 + \cdots, \label{150} \end{equation}

\noindent where $y_0$, $y_2$, $y_4 \cdots$ are constant coefficients. Substituting this series into Eq. (\ref{140}) yields

\begin{equation} 1+\left(\frac{3 y_2}{y_0}+\frac{6 y_4}{y_2}\right) z^2+\left(\frac{5 y_2^2}{y_0^2}+\frac{22 y_4}{y_0}-\frac{36 y_4^2}{y_2^2}+\frac{30 y_6}{y_2}\right) z^4+ \cdots=\kappa. \label{160} \end{equation}

\noindent Clearly, we must have $\kappa=1$, a value independent of the series coefficients. The series constants $y_0$ and $y_2$ may assume any values consistent with thermodynamic stability, but $y_4$, $y_6\cdots$ are determined by equating each coefficient on the left-hand side second-order and higher to zero, leading to $y_4= -\frac{1}{2}y_2^2/y_0$, etc. The coefficients $y_0$ and $y_2$ are two of the integration constants. The third is the first-order series coefficient $y_1$ that was set to zero in Eq. (\ref{150}). Setting $y_1=0$ results in all the odd-order series coefficients to be zero.

\par
The resulting series solution corresponds to the function:

\begin{equation} Y(z)=n_1 \sqrt{1+(n_2 z)^2},\label{170}\end{equation}

\noindent where the scaling constants $n_1$ and $n_2$ are simply related to $y_0$ and $y_2$. Direct substitution of this function into Eq. (\ref{140}) demonstrates that it is indeed the even solution analytic at $z=0$. This solution matches exactly what is known for the 1D Ising model in Eq. (\ref{80}) with $n_1=n_2=1$. This finding is the main result in this letter. An additional finding is that with $q=p$ all three of the reduced temperature expressions $t({\beta})$ in Eq. (\ref{30}) lead to the same ODE. Assuming even solutions analytic at $z=0$, all the $q=p$ cases have the solution Eq. (\ref{170}).

\par
A calculation with statistical mechanics of the scaling function $Y(z)$ in a magnetic field is usually very difficult. I am not aware of other simple worked examples not already on the list in Table 1. It is my hope that the considerations in this letter will spur such calculations.

\par
One possible candidate for further investigation is the spherical model \cite{Pathria2011, Berlin1952}, which has instances of $t(\beta)$ of the power law varieties, $2)$ and $3)$ in Eq. (\ref{30}). The spherical model was originally devised as a solvable spin model. In integer dimension $d$, it is a generalization of the Ising model. In continuous dimension $0<d<2$, it has critical point at $T_C=0$, and it has $t(\beta)=\beta^{-1}$. Baker and Bonner \cite{Baker1975} reported:

\begin{equation} \phi=\beta^{-p}\, Y\left(h\beta^{p}\right),\label{90} \end{equation}

\noindent with $p=d/(2-d)$, though no specific functional form for $Y(z)$ was given. We have $q=p$, and if we assume that $Y(z)$ is an even function analytic at $z=0$, we get the scaled form $Y(z)=\sqrt{1+z^2}$, as in Eq. (\ref{170}).

\par
The spherical model with $d>2$ has a critical point at nonzero temperature \cite{Pathria2011}. Janke et al. \cite{Janke2003} discussed this case in the context of calculating $R$, and stated that $t(\beta)=\beta_C-\beta$, putting it into category 3) in Eq. (\ref{30}). But no $Y(z)$ was reported.

\section{Conclusion}

\par
In conclusion, I have extended an information geometric method for calculating scaling functions to cases involving spin systems. I focused on the 1D ferromagnetic Ising model, and showed that the method produces the correct scaling function. Calculations of the scaling functions with information geometry are straightforward; they involve solving differential equations. Calculations with statistical mechanics in models are considerably harder, and there very few known cases in field. The challenge is to produce more.

 \end{document}